\documentclass[useAMS,usenatbib]{mn2e}
\usepackage{epsfig, nicefrac}

\title{On the emergent System Mass Function: the contest between accretion and fragmentation}
\author[]
{Paul C.~Clark \& Anthony P.~Whitworth \\
School of Physics and Astronomy, Queen's Buildings, The Parade, Cardiff University, Cardiff, CF24 3AA, UK \\
{\tt email:} clarkpc@cardiff.ac.uk; anthony.whitworth@astro.cf.ac.uk}
 
\begin{document}
\maketitle

\begin{abstract} %[213]%
We propose a new model for the evolution of a star cluster's System Mass Function (SMF). The model involves both turbulent fragmentation and competitive accretion. Turbulent fragmentation creates low-mass seed proto-systems (i.e. single and multiple protostars). Some of these low-mass seed proto-systems then grow by competitive accretion to produce the high-mass power-law tail of the SMF. Turbulent fragmentation is relatively inefficient, in the sense that the creation of low-mass seed proto-systems only consumes a fraction, $\sim\!23\%$ (at most $\sim\!50\%$), of the mass available for star formation. The remaining mass is consumed by competitive accretion. Provided the accretion rate onto a proto-system is approximately proportional to its mass ($dm/dt\!\propto\!m$), the SMF develops a power-law tail at high masses with the Salpeter slope ($\sim\!-2.3$). If the rate of supply of mass accelerates, the rate of proto-system formation also accelerates, as appears to be observed in many clusters. However, even if the rate of supply of mass decreases, or ceases and then resumes, the SMF evolves homologously, retaining the same overall shape, and the high-mass power-law tail simply extends to ever higher masses until the supply of gas runs out completely. The Chabrier SMF can be reproduced very accurately if the seed proto-systems have an approximately log-normal mass distribution with median mass $\sim\! 0.11 \,{\rm M}_{_\odot}$ and logarithmic standard deviation $\sigma_{\log_{10}(M/{\rm M}_\odot)}\sim 0.47$).
\end{abstract}

\begin{keywords}
galaxies: ISM -- ISM: clouds -- ISM: molecules --  stars: formation
\end{keywords}

%%%%
\section{Introduction}\label{sec:intro}
%%%%%

Theories for the origin of the System Mass Function (SMF), and hence also the stellar Initial Mass Function (IMF), can be divided between two main categories, which are distinguished by the scale on which most of the mass of a proto-system is accumulated. The first category comprises {\it turbulent fragmentation} theories, in which a star-forming cloud is presumed to break up into a population of prestellar cores \citep[e.g.][]{PadoanNordlund2002, HennebelleChabrier2008, HennebelleChabrier2009, Qey2011, Hopkins2012, HennebelleChabrier2013}.  The final mass of a proto-system is then determined by the local mass reservoir in a prestellar core, on relatively small scales $\la 0.1\,{\rm pc}$. The masses of cores are determined primarily by supersonic turbulence, and there is little subsequent accretion onto a proto-system from the rest of the parent cloud.  Turbulent fragmentation theories are supported by the fact that the observed prestellar core mass function (CMF) appears to have the same overall shape as the SMF and the IMF \citep[e.g.][]{Motte_etal_1998, TestiSargent1998, Johnstone_etal_2000, Johnstone_etal_2001, NutterWardThompson2007, Andre_etaL2010}, although there are some apparent exceptions to this \citep[e.g.][]{MotteFetal2018}.

The second category comprises {\it competitive accretion} theories, in which the SMF is largely determined by how an embryonic proto-system competes with other proto-systems for the mass of the parent cloud \citep{Zinnecker1982, Bonnell_etal_2001a, Bonnell_etal_2001b, Bonnell_etal_2004}. In these theories, turbulent fragmentation of the parent cloud is important in setting the peak of the SMF \citep{Larson1985, Larson05, Jappsen_etal_2005, Bonnell_etal_2006, LeeHennebelle2018, Hennebelle_etal_2019}, but the final distribution of proto-system masses above the peak is regulated by the accident of birth. Proto-systems that are born in the dense gas near the bottom of the parent cloud's gravitational potential, and are moving slowly, accrete rapidly and end up as high-mass systems. In contrast, proto-systems that are born in the more diffuse gas towards the edges of the parent cloud, and/or are moving fast, accrete much less, so their masses remain close to the peak. In the case of the high-mass proto-systems, their final mass may have been gathered from very disparate locations within the parent cloud. 

Both theories have issues. A critical issue with turbulent fragmentation theories is that, although the similarity between the shapes of CMF, SMF and IMF is suggestive, the CMF is highly uncertain, because it is very difficult to determine whether an observed core is truly prestellar (i.e. subvirial). First, we have limited information about the different energy modes in a core (e.g. \citealt{Enoch_etal_2008, LomaxWhitworth2016}). Second, the procedures used to define the boundary of a core are arbitrary and this has a strong influence on estimates of a core's virial balance (e.g. \citealt{Smith_etal_2009, GongOstriker2011}). Third, there are uncertainties associated with converting dust fluxes into core masses \citep[e.g.][]{HowardAetal2019, PriesWhitw2020}. It is therefore unclear how reliable estimated core masses are. Moreover, even if the estimated masses of prestellar cores are reliable, mapping the CMF into the SMF and then the IMF is fraught with problems \citep[e.g.][]{Clark_etal_2007, HolmanKetal2013, Offner_etal_2014}.

A critical issue with competitive accretion theories is that they postulate the existence of a population of preexisting, low-mass proto-systems, which then share the gaseous reservoir from which they accrete; this is an unrealistic starting point. Numerical simulations of cluster formation, in which the proto-systems (modelled by sink particles) form self-consistently from a turbulent cloud, have shown that this postulate can be relaxed (e.g. \citealt{Bonnell_etal_2003, Bonnell_etal_2004, Clark_etal_2008}), because fragmentation to produce new proto-systems continues after accretion onto existing proto-systems has started, and competitive accretion then regulates the high-mass end of the emerging SMF.  However the initial conditions for these simulations are still rather artificial, i.e. a very dense ($n_{_{\rm H_2}}\!\sim\!10^5 \, \rm cm^{-3}$) cloud containing many ($\sim\!1000$) Jeans masses.

It is unclear how such a heavily Jeans-unstable cloud could occur in nature,  unless there were a sudden burst of synchronised cooling.\footnote{It has been suggested that this may occur in very metal poor gas ($ 10^{-5}\la Z/Z_{_\odot}\la 10^{-3}$)  via gas-dust coupling \citep{Omukai_etal_2005, Clark_etal_2008, Dopcke_etal_2011, Dopcke_etal_2013}}  Indeed, there is growing evidence that star formation starts as soon as bound gas is assembled, and hence that star formation is concurrent with the growth of a star-forming cloud \citep{Hartmann_etaL2001, Banerjee_etal_2009,  Hartmann_etal_2012, ZA_etal_2012, Kirk_etal_2013, Peretto_etal_2013, Kruijssen_etal_2015, SmilgysBonnell2016, Barnes_etal_2018,  Urquhart_etal_2018, VS_etal_2019}. The evolution of the emerging SMF {\em may} therefore depend on both the rate at which new proto-systems form, and accretion onto existing proto-systems. 

One weakness of both turbulent fragmentation theories and competitive accretion theories is that they do not normally consider how the emerging SMF develops in time.  A comparison of the SMFs from different star-forming regions reveals that clusters of very different ages, sizes, and densities are all well described by a power law with a \citet{Salpeter1955}  slope of $\sim\!- 2.3$ at the high-mass end \citep{Bastian_etal_2010, Offner_etal_2014}.  This suggests that the shape of the emerging SMF is broadly time-invariant, at least over the observed timescales of $\ga 0.5\,{\rm Myr}$. This is an important feature of the SMF, since in a given region the star formation could be terminated at any stage by a wide variety of feedback processes, both internal \citep{RogersPittard2013, Dale_etal_2014, Dale2017, Rahner_etal_2017, Rahner_etal_2019} and external \citep{Padoan_etal_2016, Padoan_etal_2017, Seifried_etal_2018}.  Although some work has been done on the time-dependence of turbulent fragmentation \citep{HennebelleChabrier2013}, it is difficult to relate this to a growing star-forming region, in which the internal conditions are evolving. 

In this paper we consider a new model for the evolution of a star cluster's SMF. This model focuses on the balance between the formation of new proto-systems and accretion onto existing proto-systems.  The model -- which is based on the ideas presented in \citet{Dopcke_etal_2011} -- conflates the two aspects of star formation theory that are best supported by numerical simulations: (i) that the peak of the SMF is determined by turbulent fragmentation (e.g. \citealt{Jappsen_etal_2005})  and (ii) that competition for residual mass is unavoidable if the proto-systems thus formed sit within a common gravitational potential \citep{BonnellBate2006}. We therefore explore how a mixture of turbulent fragmentation {\em and} competitive accretion can deliver a time-invariant SMF.  The conversion of the SMF into the IMF is not addressed here, since this involves additional physical processes like disc fragmentation and the dissolution of multiple systems.

In Section \ref{sec:basicmodel} we derive an analytic model for the evolution of the SMF in a forming star cluster (full details are given in the Appendix) and the constraints that must be met for a power-law tail with constant slope to develop. In Section \ref{sec:sens} we explore how these constraints can be relaxed.  In Section \ref{sec:disc} we discuss the physics behind the model: the mass dependence of the accretion rate; the division of mass between the creation of seed proto-systems (by turbulent fragmentation) and competitive accretion; the acceleration of star formation in a forming star cluster; and aspects of star formation that might corrupt the model. In Section \ref{SEC:nonsteady} we show that our model still produces the same time-invariant SMF, with a Salpeter slope at high masses, when the supply of mass to the cluster varies with time -- even if, for example, the mass supply cuts off abruptly and then resumes. In Section \ref{sec:summary} we summarise our conclusions.

%%%%%
\section{An idealised analytic model}\label{sec:basicmodel}
%%%%%

In this study we are primarily concerned with the high-mass end of the SMF. If at time $t$ a cluster has $N(t,m)$ proto-systems with mass below $m$, our model requires that  
\begin{equation}\label{eq:imf}%CHECKED%
\left.\frac{\partial\!N}{\partial m}\right|_t=A(t)\,m^{-\alpha}\,.
\end{equation}
In other words, the amplitude of the SMF, $A(t)$, increases with time as more proto-systems form, but the SMF is always a power law with a constant slope $-\alpha$. In the standard formulation of the SMF, $\alpha$ takes a value of $2.3 \pm 0.3$ for masses $m\ga {\rm M}_{_\odot}$ (e.g. \citealt{Chabrier2003}).\footnote{In mathematical expressions, we use standard brackets exclusively to denote functional dependence, as for example in $A(t)$.} 

In competitive accretion theories, low-mass seed proto-systems with masses around the peak of the SMF are formed by turbulent fragmentation of a proto-cluster cloud. Some of these proto-systems then develop into high-mass proto-systems by competing successfully for the remaining cloud gas. \citet{Bonnell_etal_2001b} postulate that there should be two accretion regimes: `tidal-lobe', and `Bondi-Hoyle'. Tidal-lobe accretion is presumed to dominate when the potential of the cluster is still dominated by gas. In a spherically symmetric, centrally condensed cloud the resulting accretion rate onto a proto-system of mass $m$ is given by,
\begin{equation}%CHECKED%
\frac{dm}{dt}\propto m^{2/3}\,.
\end{equation}
In contrast, Bondi-Hoyle accretion dominates when the cluster potential is dominated by proto-systems, and the resulting accretion rate is then given by
\begin{equation}%CHECKED%
\frac{dm}{dt}\propto m^{2}\,.
\end{equation}
We will adopt a general accretion rate of the form
\begin{equation}\label{eq:mdotgen}%CHECKED%
\frac{dm}{dt}=B\,m^{\beta}\,.
\end{equation}
Provided $\beta\!>\!0$ this results in competitive accretion, since more massive proto-systems grow faster and become even more massive, i.e. ``the rich get richer''.

The high-mass end of the SMF appears to be broadly invariant with cluster mass, cluster age, and environment. Therefore in a growing cluster, the gas reservoir must continually undergo turbulent fragmentation to form new low-mass seed proto-systems, as well as accreting onto existing proto-systems to increase their masses. Otherwise the low-mass proto-systems will steadily disappear, and the SMF will become increasingly top-heavy. The SMF will then depend on when the cluster is observed, and the final SMF will depend on when accretion is terminated by exhaustion of the gas supply. 

In the Appendix, we develop an analytic model for the SMF in which new seed proto-systems are continually injected at some characteristic low mass, $m_0$, and then grow according to Equation \ref{eq:mdotgen}. We show that the amplitude of the SMF (see Equation \ref{eq:imf}) must evolve according to
\begin{eqnarray}\label{EQN:dlnAdt}%CHECKED%
\frac{d\ln(A)}{dt}&=&[\alpha -\beta]\,B\,m^{[\beta -1]}.
\end{eqnarray} 
Since $A(t)$ is by construction independent of $m$, it follows that $\beta = 1$. In other words, a necessary, but not sufficient, condition for the slope of the high-mass end of the SMF to remain exactly constant is that proto-systems grow at a rate
\begin{equation}\label{EQN:dmdt}%CHECKED%
\frac{dm}{dt} = B\,m\,;
\end{equation}
$B^{-1}$ is the e-folding time for growth of a proto-system. Values of $\beta$ above (below) unity cause the SMF to flatten (steepen) over time. We explore the effects of $\beta \ne 1$ further in Section \ref{sec:sens}.

$\beta = 1$ appears to be incompatible with both tidal-lobe accretion theory ($\beta =2/3$) and Bondi-Hoyle accretion theory ($\beta = 2$). However, the results of \citet{Maschberger_etal_2014}, in particular their Figure 6, suggest that $\beta \approx 1$ gives a good fit to the accretion rates in the simulations of cluster formation by \citet{Bonnell_etal_2011}. The physical motivation for $\beta = 1$ can still be related to Bondi-Hoyle accretion \citep{Hsu_etal_2010, BP_etal_2015, Kuznetsova_etal_2015}, and we will discuss this further in Section \ref{sec:disc}.

If we set $\beta = 1$, Equation \ref{EQN:dlnAdt} gives
\begin{equation}\label{eq:at}%CHECKED%
A(t) = A_0 \, e^{[\alpha -1]\, Bt},
\end{equation}
which together with Equation \ref{eq:imf} completely describes the evolution of the SMF. Since the first seed proto-system is introduced at $t\!=\!0$, the most massive proto-system at time $t$ has mass 
\begin{equation}\label{EQN:mMAX.1}%CHECKED%
m_{\rm max}(t) = m_0\, e^{Bt}\,.
\end{equation}
The total number of proto-systems at time $t$ is therefore
\begin{eqnarray}\nonumber%CHECKED%
{\cal N}(t)&=&A(t) \int\limits_{m=m_0}^{m=m_{\rm max}(t)} m^{-\alpha}\,dm\\\label{eq:nt}
&=&\frac{A_0 \left[ e^{[\alpha -1] Bt} - 1 \right] }{[\alpha - 1]\,m_0^{[\alpha -1]}}\,,
\end{eqnarray}
and the rate of creation of seed proto-systems is
\begin{equation}\label{eq:dndt}%CHECKED%
\frac{d{\cal N}}{dt} = \frac{A_0 \, B \, e^{[\alpha -1] B t}}{m_0^{[\alpha -1]}}\,.
\end{equation}
Thus the necessary and sufficient condition for the slope of the high-mass end of the SMF to remain constant is that turbulent fragmentation creates low-mass seed proto-systems at the rate given by Equation \ref{eq:dndt}, and that these proto-systems then accrete according to Equation \ref{EQN:dmdt}.

It follows that the total mass of the cluster at time $t$ is
\begin{eqnarray}\nonumber%CHECKED%
M_{\rm tot}(t) &=& A(t) \int\limits_{m=m_0}^{m=m_{\rm max}(t)} m^{1-\alpha}\,dm \\\label{eq:mtotal}
&=& \frac{ A_0 \left[ e^{[\alpha -1]Bt} - e^{Bt} \right] }{\left[ \alpha - 2 \right]\,m_0^{[\alpha -2]}}\,,
\end{eqnarray}
and hence the rate at which matter is consumed by star formation is
\begin{eqnarray}\label{mdottotal}%CHECKED%
\frac{dM_{\rm tot}}{dt} &=&\frac{ A_0 B \left[ [\alpha -1] e^{[\alpha -1]Bt} - e^{Bt} \right] }{\left[ \alpha - 2 \right]m_0^{[\alpha -2]}}\,.
\end{eqnarray}
This is divided between the rate at which mass is consumed by turbulent fragmentation ({\sc tf}) creating low-mass seed proto-systems,
\begin{eqnarray}\label{eq:dmseeddt}%CHECKED%
\frac{dM_{_{\rm TF}}}{dt}&=&\frac{d{\cal N}}{dt}\,m_0\;\,=\;\,\frac{ A_0 B e^{[\alpha -1]Bt}}{m_0^{[\alpha -2]}}\,,
\end{eqnarray}
and the rate at which mass is consumed building higher-mass proto-systems by competitive accretion ({\sc ca}),
\begin{eqnarray}%CHECKED%
\frac{dM_{_{\rm CA}}}{dt}&=&BM_{\rm tot}(t)\;\,=\;\,\frac{ A_0 B \left[ e^{[\alpha -1]Bt} - e^{Bt} \right] }{\left[ \alpha - 2 \right]m_0^{[\alpha -2]}}.
\end{eqnarray}

Hence the fraction of the consumed mass that goes into forming seed proto-systems by turbulent fragmentation is
\begin{eqnarray}\nonumber%CHECKRD%
f(t)&=&\frac{dM_{_{\rm TF}}/dt}{dM_{\rm tot}/dt)}\\\label{eq:massfractime}
&=&\frac{[\alpha -2]}{[\alpha -1]} \left\{1 - \frac{e^{-[\alpha -2]Bt }}{[\alpha -1]} \right\}^{\!-1}.
\end{eqnarray}
In the limit $t \gg B^{-1}$ (i.e., with the value of $B$ that we will adopt below, $t \gg 10^5 \rm \, yr$), this tends to the asymptotic expression,
\begin{eqnarray}\label{eq:seedfrac}%CHECKED%
f_{_{\rm TF}}&=&\frac{[\alpha -2]}{[\alpha -1]}\,.
\end{eqnarray}
For $\alpha\!=\!2.3$ this gives $f_{_{\rm TF}}\!\simeq\!0.23$. Thus, roughly 23\% of the mass consumed forms low-mass seed proto-systems by turbulent fragmentation, and the remaining 77\% goes towards increasing the masses of existing proto-systems by competitive  accretion.  We discuss possible physical reasons for this division of mass in Section \ref{subsec:fraction}.

%%%%%
\begin{figure}
\centerline{ \includegraphics[width=3.4in]{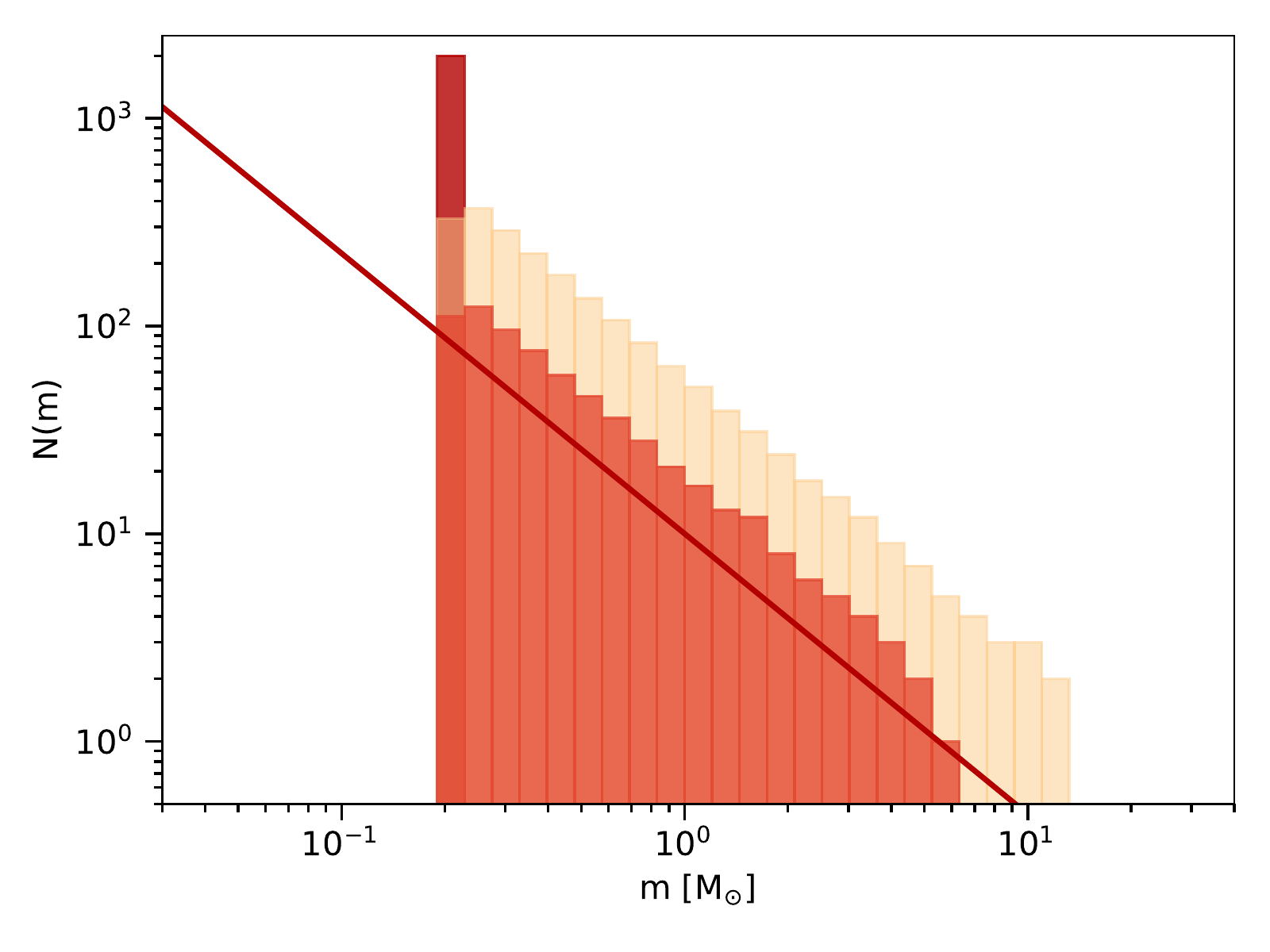} }
\caption{The evolution of the SMF for the simplest realisation of the stochastic fragmentation/accretion model. All seed proto-systems are born with $m_0\!=\!0.2\,{\rm M}_{_\odot}$, at a rate given by Equation \ref{eq:dndt} with $A_0\!=\!1$, $\alpha \!=\!2.3$, and $B\!=\! 10^{-5}\rm\,yr^{-1}$. Proto-systems then accrete mass according to Equation \ref{EQN:dmdt}.  The initial proto-system masses are shown in dark red, while the (progressively lighter) orange and yellow distributions show the emerging SMF after, respectively, 667 and 2000 proto-systems have formed. We see that the slope is approximately time-invariant, and equal to the desired $-\alpha \!=\!-2.3$, as indicated by the solid red line.}
\label{fig:fiducial}
\end{figure}
%%%%%

To test the analytic model, we construct a numerical model for the evolution of the SMF based on Equations \ref{eq:dndt} and \ref{EQN:dmdt}. Since the term $A_0 m_0^{1-\alpha}$ in Equation \ref{eq:dndt} is simply a scaling factor for the amplitude of the SMF, we set it to unity. Motivated by the results of Maschberger et al. (2014), we set $B\!=\! 10^{-5}\rm\,yr^{-1}$, but again $B$ is simply a scaling-factor for the timescale on which accretion occurs. We set $\alpha\!=\!2.3$ since this is the observed slope of the high-mass end of the SMF \citep[e.g.][]{Salpeter1955, Kroupa2001, Chabrier2003}. Finally, we set $m_0\!=\!0.2\,{\rm M}_{_\odot}$, since this is the peak of the observed SMF \citep{Chabrier2003}, and we assume that all proto-systems are born with exactly this mass. Fig. \ref{fig:fiducial} shows the initial seed proto-system mass distribution (essentially a delta-function at $m_0\!=\!0.2\,{\rm M}_{_\odot}$), and the SMFs once 667 proto-systems have formed, and once 2000 proto-systems have formed. On a log-log plot, both SMFs approximate well to the slope $-\alpha \!=\!-2.3$. The formation of new low-mass seed proto-systems by turbulent fragmentation (Equation \ref{eq:dndt}) perfectly balances the growth of proto-systems by competitive accretion (Equation \ref{EQN:dmdt}).

%%%%%
\begin{figure}
\centerline{ \includegraphics[width=3.4in]{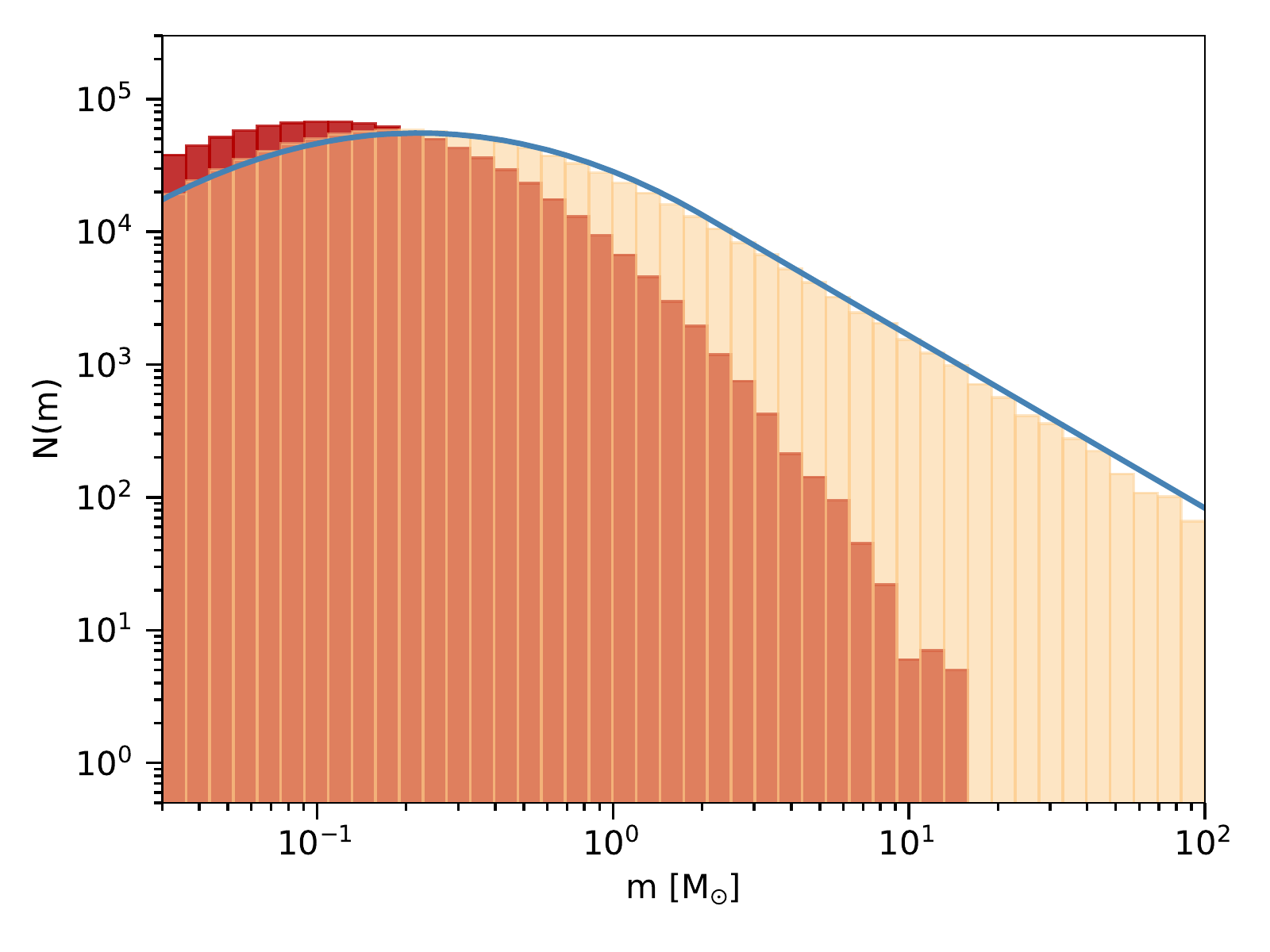} }
\caption{As Fig. \ref{fig:fiducial}, except that (i) the masses of seed (red histogram) are drawn randomly from a log-normal distribution with mean $\mu_{_{\log_{10}(M/{\rm M}_\odot)}}\!=\!-0.975$ and standard deviation $\sigma_{_{\log_{10}(M/{\rm M}_\odot)}}\!=\!0.470$, and (ii) we have generated $10^6$ proto-systems in order to improve the statistics on the SMF (yellow histogram). The blue solid curve shows the Chabrier (2003) SMF, and -- apart from low masses, $\la 0.03\,{\rm M}_{_\odot}$, where disc fragmentation and ejection are presumed to generate additional proto-systems -- the fit is extremely accurate, and well within the uncertainties.}
\label{fig:lognorm}
\end{figure}

%%%%%
\section{A more realistic stochastic model}\label{sec:sens}
%%%%%

In this section we explore how the evolution of the SMF changes when the rather precise conditions of the idealised analytic model (Equations \ref{eq:dndt} and \ref{EQN:dmdt}) are relaxed.

%%%%%
\subsection{Turbulent fragmentation}\label{SEC:TurbFrag}
%%%%%

Turbulent fragmentation is not expected to deliver a single seed proto-system mass, $m_0$. Therefore hereafter we draw seed-masses randomly from a log-normal distribution.  The parameters of this distribution (its mean and standard deviation) have been set by picking seed-masses from a trial log-normal distribution, evolving them in the same way as described in Section \ref{sec:basicmodel} and seeking a close match to the observed SMF.

For the observed SMF we take the prescription in Chabrier (2003), viz. a log-normal peak with mean $\mu_{_{\log_{10}(M/{\rm M}_\odot)}}\!=\!-0.658$ and standard deviation $\sigma_{_{\log_{10}(M/{\rm M}_\odot)}}\!=\!0.570$, and we join this smoothly to a power-law tail with $\alpha\!=\!2.3$. In order to obtain a smooth join, the switch from log-normal to power-law is at $2.03\,{\rm M}_{_\odot}$. This SMF is illustrated by the smooth solid curve on Figure \ref{fig:lognorm}.

The log-normal seed-mass distribution that best reproduces this SMF has mean $\mu_{_{\log_{10}(M/{\rm M}_\odot)}}\!=\!-0.975$ and standard deviation $\sigma_{_{\log_{10}(M/{\rm M}_\odot)}}\!=\!0.470$; the median seed-mass is therefore $m_0\!=\!0.106\,{\rm M}_{_\odot}$, and the FWHM of the seed-mass distribution extends from $0.03\,{\rm M}_{_\odot}$ to $0.38\,{\rm M}_{_\odot}$. This seed-mass distribution is represented by the red histogram on Figure \ref{fig:lognorm}, and it is adopted for all the cases discussed in the sequel.

The SMFs derived from this seed-mass distribution for 1,000,000 proto-systems is represented by the yellow histogram on Figure \ref{fig:lognorm} and this is termed the fiducial case. We see that, between $\sim\!0.03\,{\rm M}_{_\odot}$ and $\sim\!100\,{\rm M}_{_\odot}$ it is an extremely good fit to the observed SMF. Below $\sim\!0.03\,{\rm M}_{_\odot}$, the model SMF falls below the observed one, but this is the region where we expect a significant fraction of proto-systems to have been formed by other processes.

The dominant formation mechanism for very low-mass proto-systems (i.e. free-floating Brown Dwarfs and planets) is contentious (Whitworth et al. 2007). Some authors argue that Brown Dwarfs form like low-mass stars by turbulent fragmentation (e.g. Padoan \& Nordlund, 2004), but there are problems with this paradigm. In particular, it seems to requires very supersonic, and unrealistically focussed radial inflows to produce a gravitationally unstable core of Brown-Dwarf mass \citep{LomaxWhitworth2016}. The main alternative is that, as one considers lower and lower masses, an increasing proportion of stars are formed by disc fragmentation (e.g. Whitworth \& Stamatellos, 2006). Brown Dwarfs and planets formed in this way can subsequently be ejected from their birth-disc to produce a diaspora of free-floating low-mass systems. Since neither the detailed dynamics of turbulent fragmentation, nor the detailed dynamics of disc fragmentation and dynamical ejection, are addressed in this paper, we do not pursue the issue of forming very low-mass proto-systems further here.

We note that, if the best fit to the SMF changes in future with better observations, provided that it still involves a log-normal peak merging smoothly with a power-law tail at high masses, it can be fit equally well simply by adjusting the parameters of the log-normal distribution of seed-masses and the value of $\alpha$. Based on the Chabrier (2003) SMF, the log-normal seed-mass distribution has mean $\mu_{_{\log_{10}(M/{\rm M}_\odot)}}\!=\!-0.975$ and standard deviation $\sigma_{_{\log_{10}(M/{\rm M}_\odot)}}\!=\!0.470$.

%%%%%
\subsection{Competitive accretion}\label{SEC:CompAccr}
%%%%%

\begin{figure*}
\centerline{ 
	\includegraphics[width=3.5in]{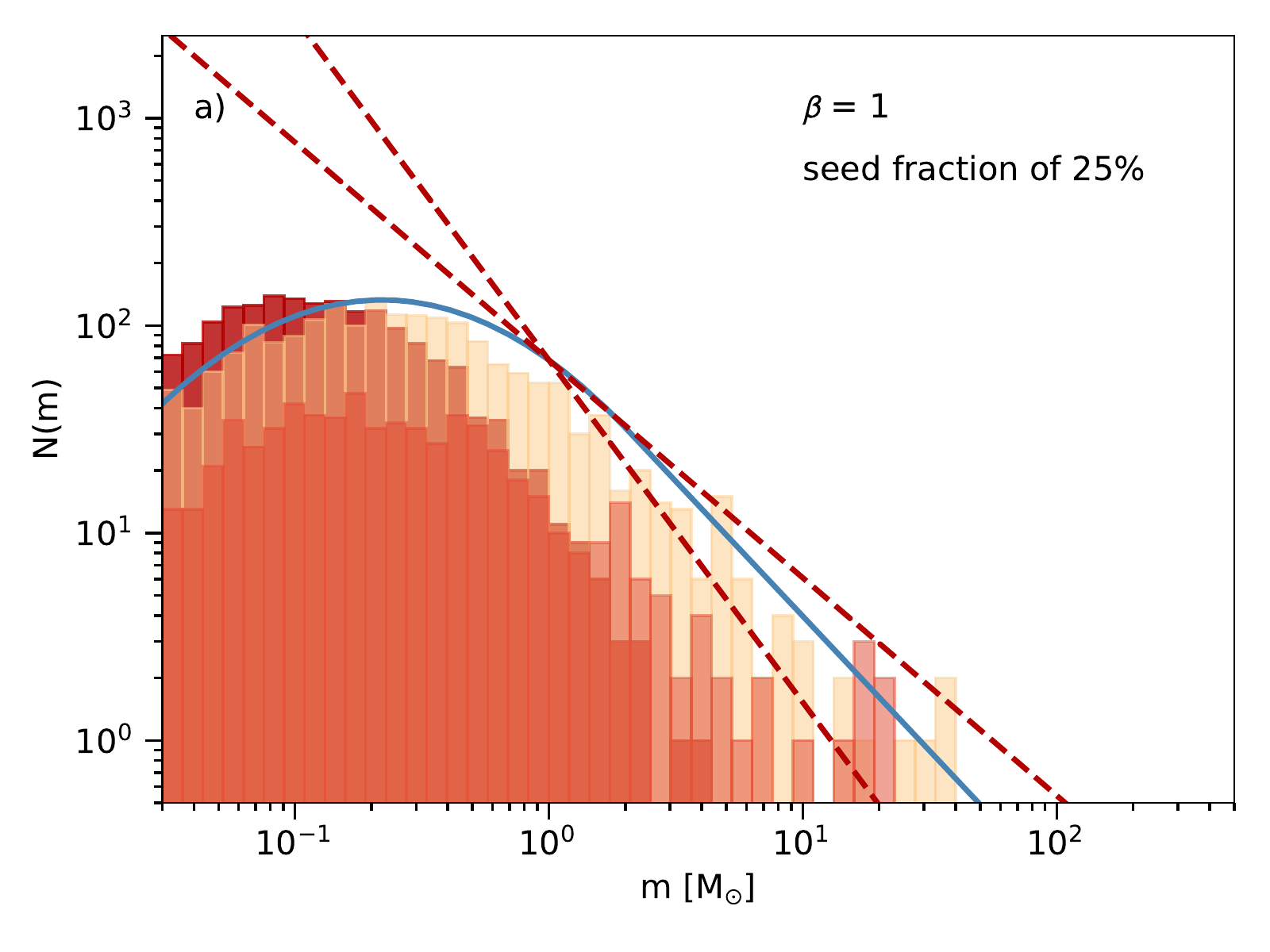} 
	\includegraphics[width=3.5in]{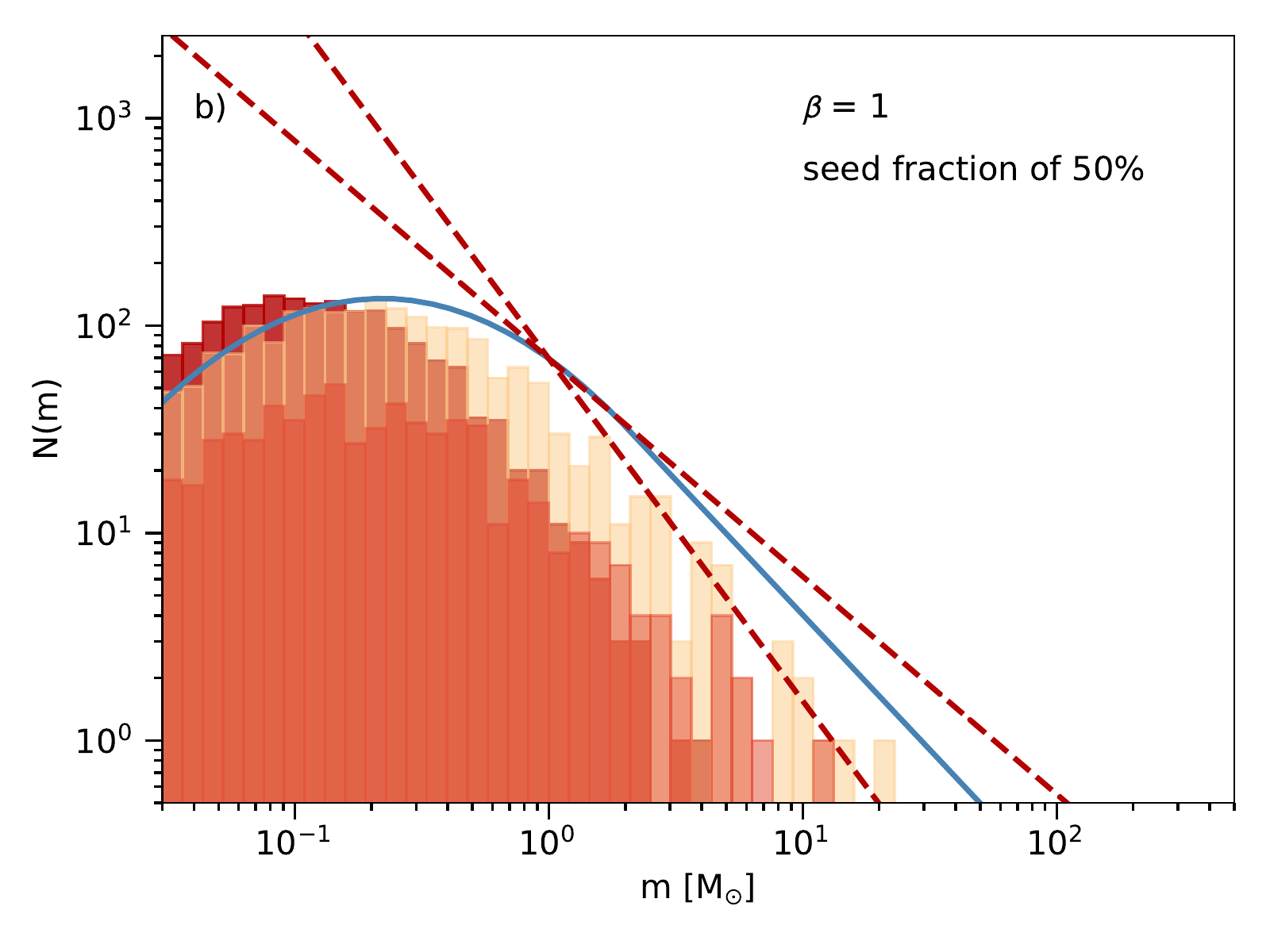} 
	}
\centerline{ 
	\includegraphics[width=3.5in]{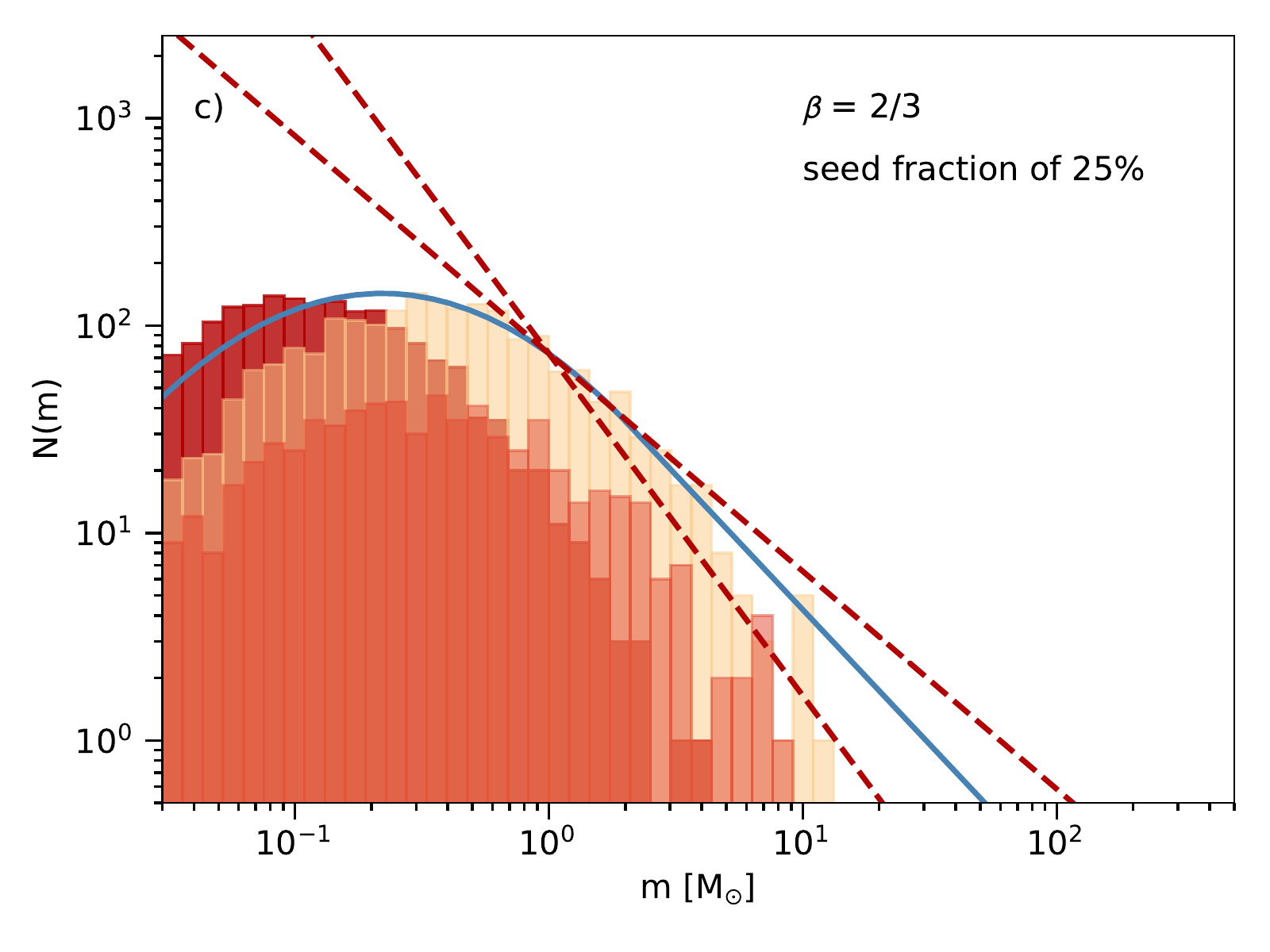} 
	\includegraphics[width=3.5in]{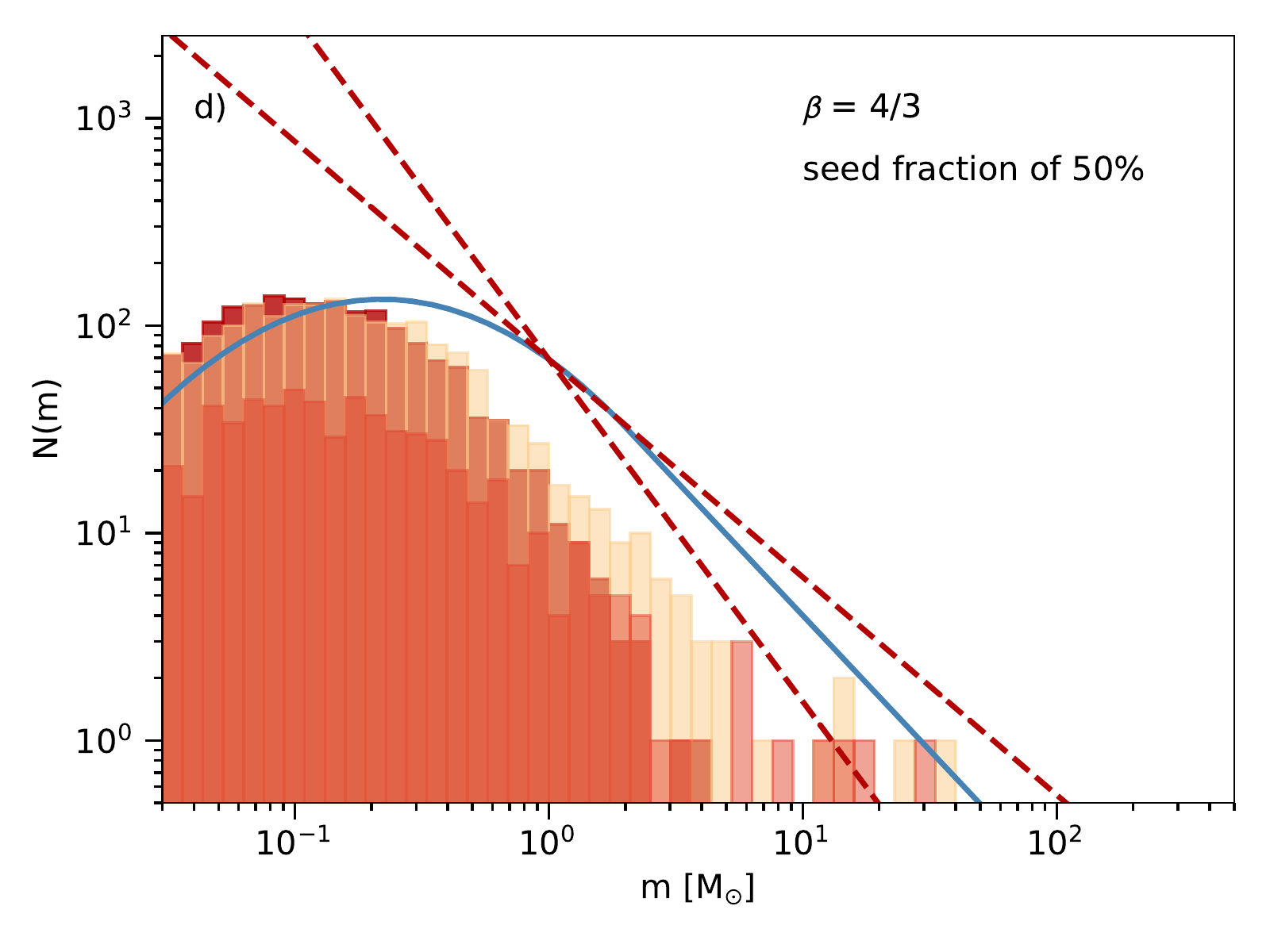} 
	}	
\centerline{ 
	\includegraphics[width=3.5in]{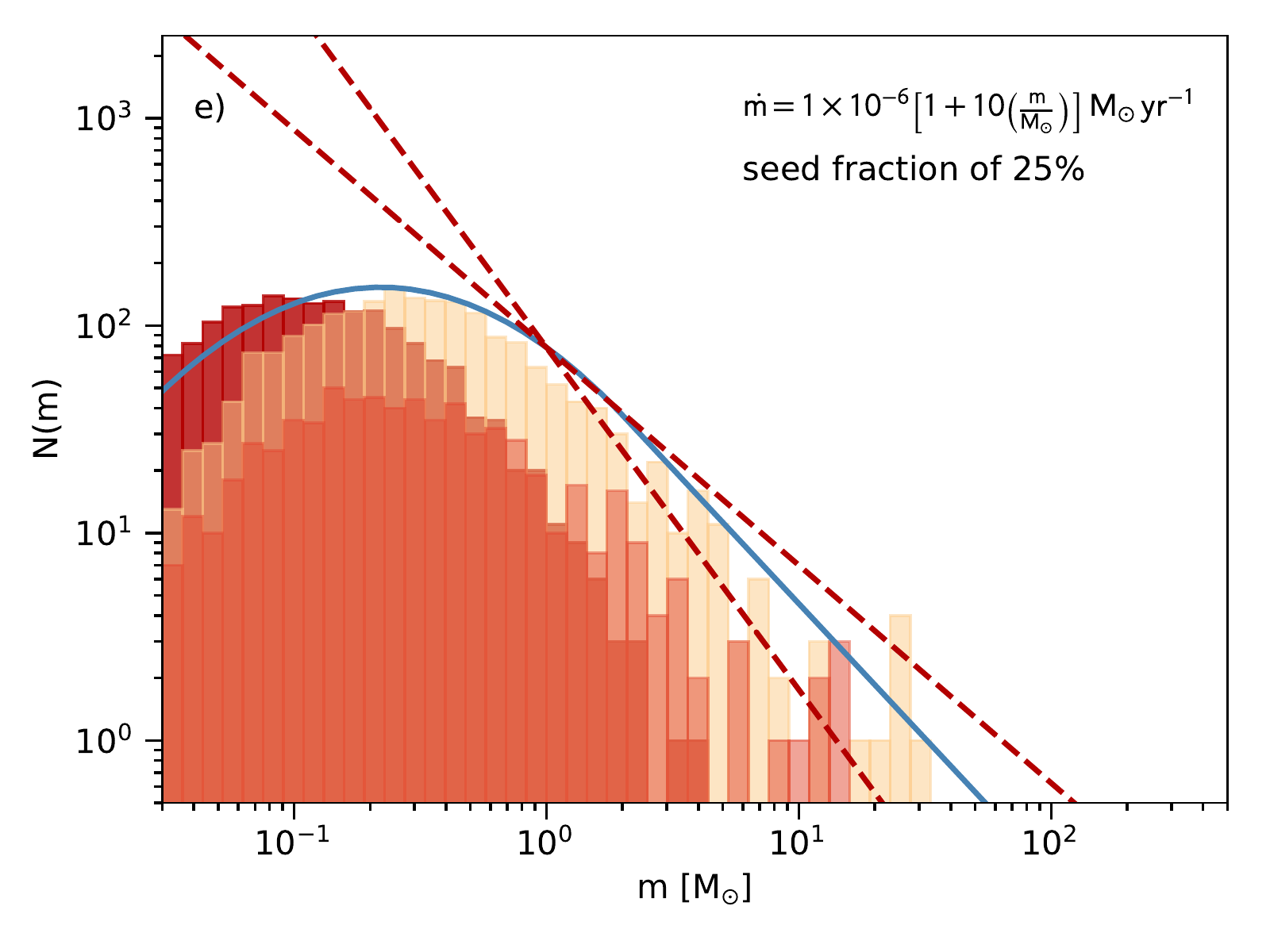} 
	\includegraphics[width=3.5in]{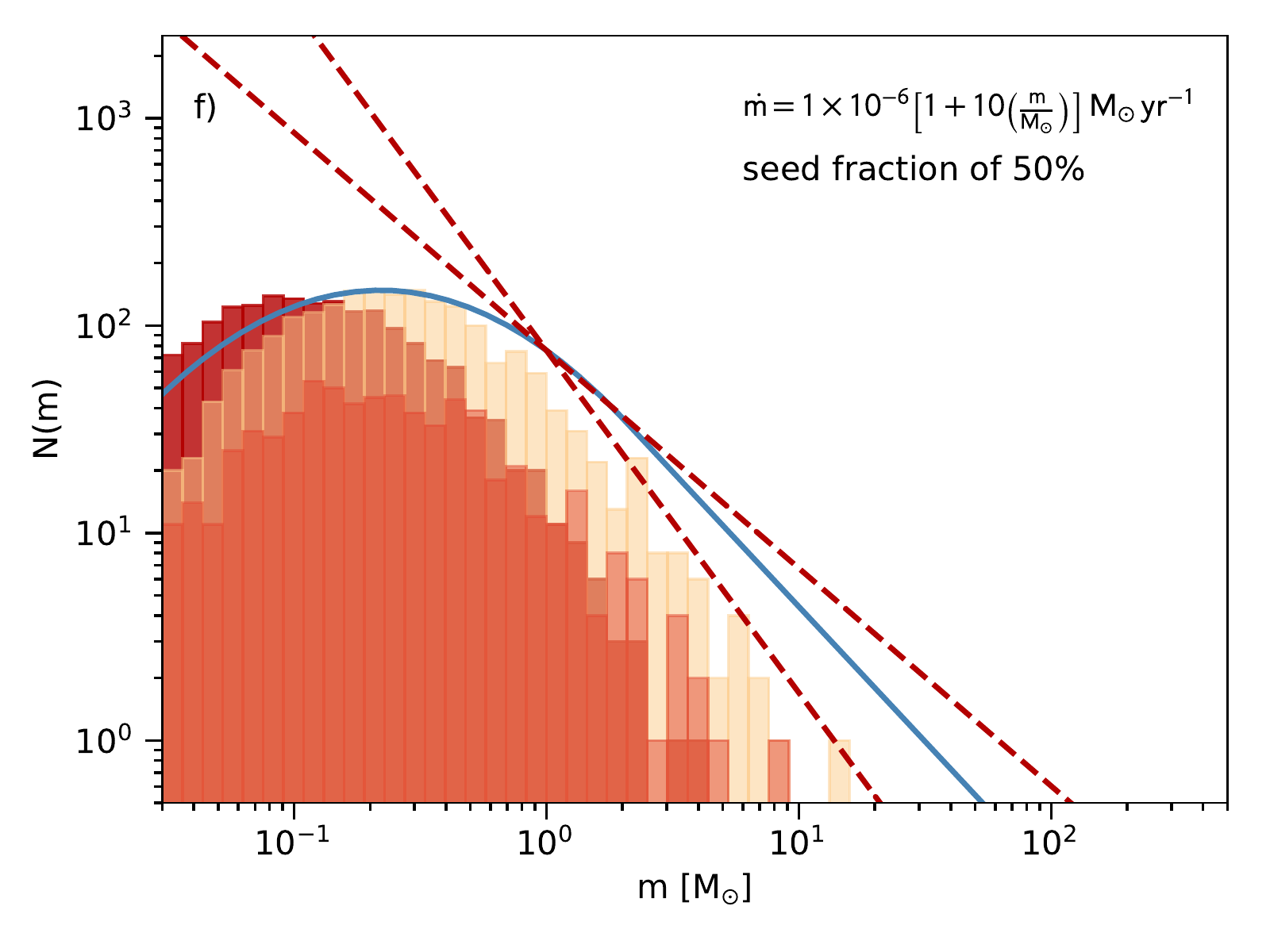} 
	}
\caption{SMFs obtained using different prescriptions for competitive accretion. On each panel, the solid red line has slope $-\alpha\!=\!-2.3$, and the dashed red lines have slopes $-\alpha\!=\!-2.0$ and $-2.6$. The initial proto-system masses are shown in dark red, while the (progressively lighter) orange and yellow distributions show the emerging SMF after 667 and 2000 proto-systems have been formed, respectively. Panels a) through d) show results obtained using different combinations of $\beta$ (Equation \ref{eq:mdotgen}) and $f_{_{\rm TF}}$. Panels e) and f) show the results obtained using Equation \ref{eq:maschdmdt} for the rate of accretion and different values of $f_{_{\rm TF}}$. One again, the blue solid curve shows the Chabrier (2003) SMF, and the red dashed lines show the uncertainties in the observed slope of the power-law portion of the SMF.}
\label{fig:sens}
\end{figure*}

In this section we explore the effects of changing the prescription for competitive accretion. First, we abandon the use of Equation \ref{eq:dndt} to regulate the rate of creation of seed proto-systems. We continue to set $A_0m_0^{[1-\alpha]}=1$, and $B\!=\!10^{-5}\,{\rm yr}^{-1}$, but we characterise the birth and growth of proto-systems by specifying the fraction of mass, $f_{_{\rm TF}}$, that goes into new seed proto-systems; values $f_{_{\rm TF}}\!=\!0.25$ and $0.50$ are treated. Note that at early times, our model actually requires $f_{_{\rm TF}}$ to be closer to unity, which means there should be an initial burst of star formation.  To mimic this effect we start our simulations with 3 systems initially.  Second, we explore variations in the exponent $\beta$ characterising the rate of mass accretion (see Equation \ref{eq:mdotgen}); values of $\beta\!=\!2/3,\;1\;{\rm and}\;4/3$ are treated. Third, we explore a modified expression for the mass-dependence of the accretion rate,
\begin{equation}\label{eq:maschdmdt}
\frac{dm}{dt} = B \left\{m + 0.1{\rm M}_{_\odot}\right\}\,;
\end{equation}
this expression has the merit that it fits better the results obtained by Maschberger et al. (2004) on the basis of a full hydrodynamical simulation; for low-mass proto-systems ($m\la 0.1\,{\rm M}_{_\odot}$), the accretion rate tends to a fixed value. The results obtained with these changes are displayed in the panels of Fig. \ref{fig:sens}. In all cases the histograms display the seed-mass distribution (as derived in Section \ref{SEC:TurbFrag}), and the SMFs after 667 and 2000 proto-systems have formed.

%%%%%
%%%%%

Fig. \ref{fig:sens}a shows the results obtained with $\beta\!=\!1$ and $f_{_{\rm TF}}\!=\!0.25$. The emergent SMF is a good fit to the observed SFM, although there is clearly some noise from the stochasticity of the model.  Fig. \ref{fig:sens}b shows the results obtained with $\beta\!=\!1$ and $f_{_{\rm TF}}\!=\!0.50$. Again the results are close to the observed SFM, but transition from log-normal to Salpeter power-law occurs a lower mass than in the Charbrier SMF, due to the increased rate of seed formation with relative to accretion.  However, given the uncertainties in the observed mass function, both these models are consistent with SMFs of real stars.

Fig. \ref{fig:sens}c shows the results obtained with $\beta\!=\!2/3$ (as appropriate for tidal-lobe accretion) and $f_{_{\rm TF}}\!=\!0.25$. With such a low $\beta$, and hence only moderately competitive accretion, lower-mass proto-systems accrete almost as fast as higher-mass ones. Consequently there is a higher proportion of  intermediate-mass proto-systems than in other cases, and a clear deficit of high-mass stars. However, for the mass range $1\, {\rm M}_{\odot} < {\rm m} < 10 \,{\rm M}_{\odot}$, there is still a good fit to the power-law portion of the observed SMF.

Fig. \ref{fig:sens}d shows the results obtained with $\beta\!=\!4/3$ and $f_{_{\rm TF}}\!=\!0.50$. With such a high $\beta$, and hence extremely competitive accretion, the first exceptionally high-mass seed proto-system to form quickly consumes a disproportionate  fraction of the available mass.  There is little time for the other proto-systems to grow much, and so their mass function is quite close to the seed-mass distribution, with a small tail on the high-mass side. We conclude that such a high value of $\beta$ is incompatible with the observed high-mass slope ($\alpha\!\simeq\!2.3$).

Figs. \ref{fig:sens}e and \ref{fig:sens}f show the results obtained when accretion is regulated by Equation \ref{eq:maschdmdt} with $f_{_{\rm TF}}\!=\!0.25\;{\rm and}\;0.50$, respectively. Above $0.3 \, \rm M_{\odot}$, the functions obtained are very similar to those obtained using $dm/dt\!=\!Bm$ with, respectively,  $f_{_{\rm TF}}\!=\!0.25$ (Fig. \ref{fig:sens}a) and $f_{_{\rm TF}}\!=\!0.50$ (Fig. \ref{fig:sens}b), although there is a slight steeping on the power-law portion of the SMF in the case with $f_{_{\rm TF}}\!=\!0.50$ in panel e in comparison to panel b.  At the low-mass end we see a larger difference between the modules of the accretion rates: the extra boost in accretion to the low-mass objects in the models following the prescription given in Equation \ref{eq:maschdmdt} results in fewer objects below $0.3 \, \rm M_{\odot}$ than the models adopting $dm/dt\!=\!Bm$.  Although this renders the models following Equation \ref{eq:maschdmdt} inconsistent with the observational data, it is worth stressing that there is a large scatter in the data in \citep{Maschberger_etal_2014} which is not captured here, and that the data in that paper is for individual stars, not systems. 

In summary, the model parameters can vary slightly, i.e. $2/3\!\la\!\beta\!\la\!1$ and $0.25\!\la\!f_{_{\rm TF}}\!\la\!0.50$, and still be broadly consistent with the observed SMF. The limits on $\beta$ suggest that tidal-lobe accretion might be a better model than Bondi-Hoyle accretion, and we return to this below.

%%%%%
\section{Discussion}\label{sec:disc}
%%%%%

The theoretical arguments outlined in the preceding sections place constraints on the evolution of the SMF in a forming star cluster, in the situation where a proto-system once formed, can grow by accretion from a common reservoir of residual gas.  In this section we discuss whether these constraints are plausible, and explore the underlying physics.

%%%%%
\subsection{The mass dependence of the  accretion rate}
%%%%%

Maschberger et al. (2014) argue that neither tidal-lobe accretion, nor  Bondi-Hoyle accretion, operates in simulations of star formation in turbulent clouds, since they find $\beta\!\sim\!1$, rather than $\beta\!\simeq\!2$.  However, their reasoning may be too simplistic. The Bondi-Hoyle accretion rate is given by
\begin{equation}\label{EQN:dmdtBH}
\frac{dm}{dt} = \frac{4\pi G^2m^2\rho_{\rm bg} v_{\rm rel}} {(v^{2}_{\rm rel}+c_{\rm s}{^2})^2 }\,.
\end{equation}
If the undisturbed background gas density, $\rho_{\rm bg}$, the velocity of a proto-system relative to this gas, $v_{\rm rel}$, and the sound speed in this gas, $c_{\rm s}$, are all approximately independent of $m$, then Equation \ref{EQN:dmdtBH} implies $\beta\!\sim\!2$. However, dynamical collapse, and exchange of energy between gas and proto-systems, produce variations in $v_{\rm rel}$. More massive proto-systems tend to be concentrated near the centre of the cluster-forming cloud, where $\rho_{\rm bg}$ is higher. Consequently there might be some extra dependence on mass that is not accounted for by $\beta\!\simeq\!2$.

\citet{BP_etal_2015} find that this is indeed the case. In different regions of their simulation, accretion subscribes locally to the Bondi-Hoyle rate, i.e.  $dm/dt\propto m^2\rho_{\rm bg}/v_{\rm rel}^3$ (since $v_{\rm rel}\!\gg\! c_{\rm s}$).  \citet{Kuznetsova_etal_2018} find the same result (see their Fig. 11).  At the same time, they also recover the Maschberger et al. (2014) result that $dm/dt\!\propto\!m$, because higher-mass proto-systems tend to be concentrated in regions where $\rho_{\rm bg}/v_{\rm rel}^3$ is lower. They conclude that Bondi-Hoyle accretion is the underlying mechanism for proto-system growth, but variations in $\rho_{\rm bg}/v_{\rm rel}^3$ are anti-correlated with $m$ in such a way that $\beta\!\sim\!1$.  

It remains to be understood why $\rho_{\rm bg}/v_{\rm rel}^3$ is anti-correlated with $m$ in this way. Where higher-mass proto-systems are forming, the density must be lower, and/or the velocity dispersion must be higher. One possibility is that dynamical stirring of the gas by more massive proto-systems increases the velocity dispersion, and this in turn reduces the gas density.

Alternatively, since we have shown that an acceptable SMF can be produced with $\beta\!=\!2/3$ (Fig. \ref{fig:sens}c), and given the large scatter in the simulations \citep{Maschberger_etal_2014, BP_etal_2015, Kuznetsova_etal_2015, Kuznetsova_etal_2018}, it may be that in nature tidal-lobe accretion dominates over Bondi-Hoyle accretion.

%%%%%
\subsection{The fraction of mass going into seeds}\label{subsec:fraction}
%%%%%

A key element of our model is that turbulent fragmentation is rather inefficient, in the sense that only a fraction, $f_{_{\rm TF}}$, of the mass that goes into forming a star cluster is used to form new low-mass proto-systems by turbulent fragmentation, and the rest is accreted onto existing proto-systems.  While inefficient fragmentation is seen in many cluster formation simulations (e.g. \citealt{Clark_etal_2008, Bonnell_etal_2008, Offner_etal_2008, Offner_etal_2009, Girichidis_etal_2011, FederrathKlessen2012, FederrathKlessen2013}), it is unclear how physics conspires to deliver the fraction $0.23\!\la\!f_{_{\rm TF}}\!\la\!0.50$ required by our model (see Section \ref{SEC:CompAccr}).

Inefficient fragmentation may arise in part due to the non-homologous nature of gravitational collapse. Prestellar cores develop strong central condensations as they collapse \citep{BodenheimerSweigart1968, Larson1969, Penston1969a, Penston1969b, ShuFH1977, WhitwSumme1985}. For an isothermal core, the outer parts tend towards a density profile $\rho (r) \propto r^{-2}$, and fall towards the centre of the core relatively slowly. This means that most of the mass of a core is still quite diffuse when the centre of the core undergoes fragmentation \citep{Lomax_etal_2015, LomaxWhitworth2016}. For a prestellar core that starts collapse at density $n_{_{\rm H_2}}\sim10^{5}\,{\rm cm^{-3}}$, only about 1 percent of the mass is involved in the buildup of the first, central, optically thick seed proto-system \citep{Larson1969}. Much of the remaining core mass is at this stage still quite far from the central seed proto-system and only moving inwards slowly.

If the core is part of a forming star cluster (i.e. a proto-cluster), and bound to the proto-cluster, the core will tend to freefall towards the centre of the proto-cluster. If the mean density of gas and stars closer to the centre of the proto-cluster is ${\bar\rho}_{\rm clust}$, any core gas with density less than $\sim\!2{\bar\rho}_{\rm clust}$ is likely to be tidally stripped from the core.  Since most of the mass in a collapsing core remains at low density for at least one core free-fall time ($t_{\rm core}\sim[G\rho_{\rm core}]^{-1/2}$, typically $\ga\!10^5\,{\rm yr}$), this low-density gas can be stripped from the core, rather than accreting onto the central proto-system. There is therefore a competition between the time it takes for the outer layers of a core to collapse onto the proto-system at the centre of the core, and the time it takes for a core formed in the outer reaches of a proto-cluster cloud to fall towards the centre of the proto-cluster cloud. Once the core reaches the centre of the proto-cluster, it can grow further by gravitational accretion, but it is now in competition with other proto-systems. This process was first explored by \citet{Bonnell_etal_2008}, where it was proposed as the mechanism by which clusters form low-mass stars and brown dwarfs.

Finally, the formation of prestellar cores by supersonic turbulence is intrinsically an inefficient process. Provided the mass supplied to a forming cluster is turbulent, the rate of formation of seed proto-systems will be slow \citep{Smith_etal_2009}.

%%%%%
\begin{figure}
\vspace{-1.2cm}\hspace{0.2cm}
\centerline{ \includegraphics[width=0.92\columnwidth,angle=270]{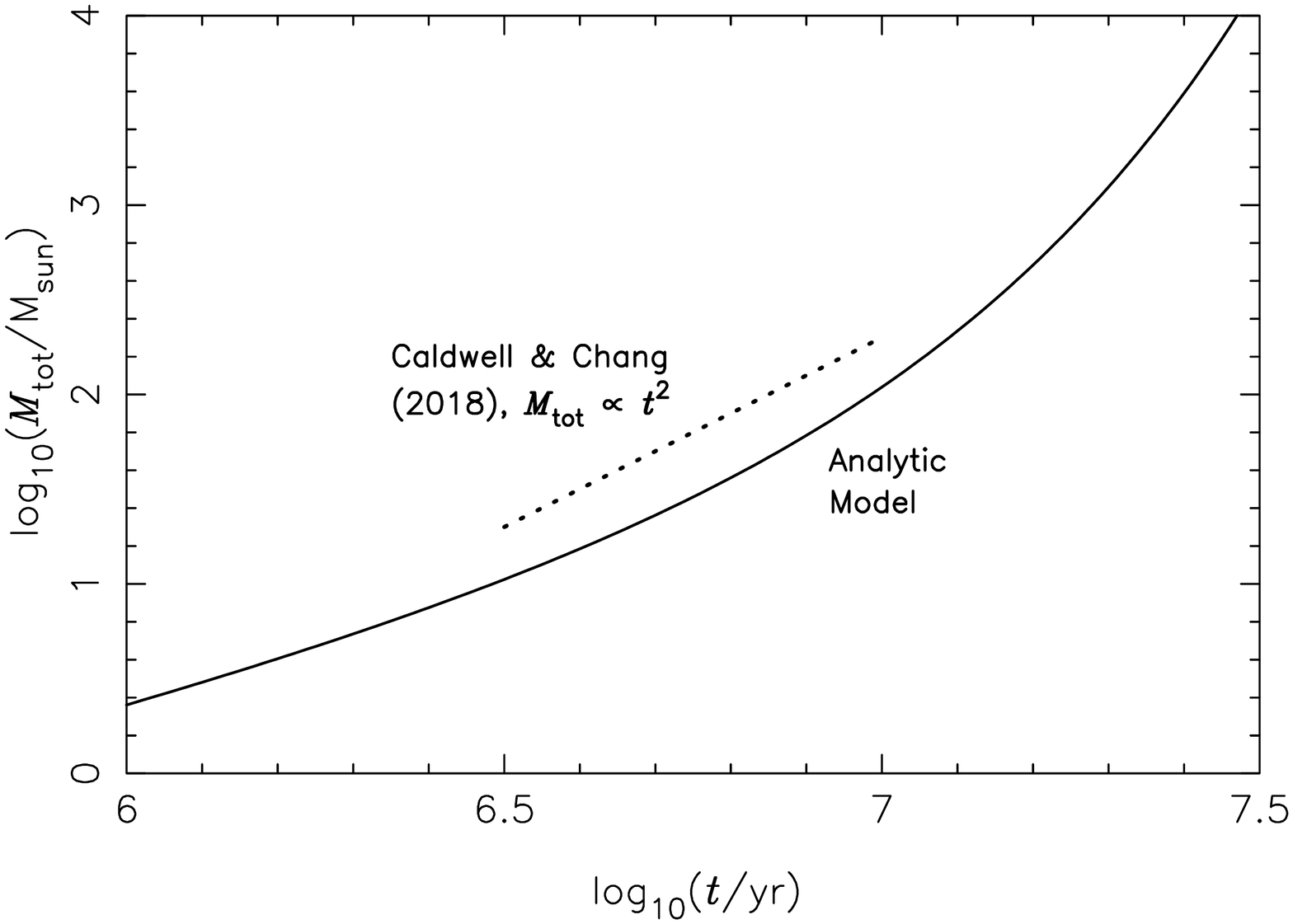} }
\caption{The total mass of proto-systems, $M_{\rm tot}$, as predicted by our analytic model (full line, Equation \ref{eq:mtotal} with $A_0=7.96\,{\rm M}_{_\odot}^{1.3}$ and $B=1.53\times 10^{-7}\,{\rm yr}^{-1}$) and by the fitting formula of Caldwell \& Chang (dotted line, Equation \ref{EQN:C&CFit}). The Caldwell \& Chang fit is only made over half a dex (as are all their fits) and has been offset by $+\!\log_{_{10}}\!(2)\!=\!+0.301$ to avoid confusion. Without this offset the two curves agree to within $\pm 10\%$ between $3\times 10^6\,{\rm yr}$ and $10^7\,{\rm yr}$.}
\label{fig:clustergrowth}
\end{figure}
%%%%%

%%%%%
\subsection{Accelerating star formation}
%%%%%

Our model requires that the rate of production of seed proto-systems grow exponentially with time, according to Equation \ref{eq:dmseeddt}. An accelerating star formation rate has been inferred from observations of nearby star-forming regions \citep{PallaStahler1999, PallaStahler2000, PallaStahler2002} and forms the basis of several theories of cloud and cluster assembly \citep[e.g.][]{Hartmann_etaL2001, Hartmann_etal_2012, ZA_etal_2012,  MurrayChang2015, VS_etal_2017, VS_etal_2019}.

\citet{CaldwellChang2018} have performed a new analysis of four nearby star-forming regions. Based on pre-main-sequence age estimates from the literature, they conclude that the star formation rates in these regions are accelerating, and they fit their results with a relation of the form $M_{\rm tot}(t)\propto t^\gamma$ with $\gamma\!\sim\!2$. These fits are obtained over a rather small time range, typically half a dex, and \citet{CaldwellChang2018} do not discuss their uncertainties, so the fits are only indicative. Our analytic model predicts a different fit function (Equation \ref{eq:mtotal}), but, given the limitations of the fitting process, it is compatible with the \citet{CaldwellChang2018} data, as we show in Fig. \ref{fig:clustergrowth}.

We note that our choice of $B=10^{-5}\,{\rm yr}^{-1}$ is motivated purely by the desire to match the simulations of \citet{Bonnell_etal_2011} and the analysis of those simulations by \citet{Maschberger_etal_2014}. The \citet{Bonnell_etal_2011} simulations start with a very dense cloud, ${\bar n}_{_{\rm H_2}}\sim 10^5\,{\rm cm^{-3}}$, and therefore evolve on a very short freefall timescale, $\sim 10^5\,{\rm yr}$. Consequently our model evolves on a comparably short timescale: 2000 proto-systems take $\sim 0.5\,{\rm Myr}$ to form. Nearby star forming clouds appear to have significantly lower mean densities, ${\bar n}_{_{\rm H_2}}\la 100\,{\rm cm^{-3}}$ and hence significantly longer freefall timescales, $\ga 3\times 10^6\,{\rm yr}$. Therefore in order to compare our results with \citet{CaldwellChang2018} we must stretch the time axis. This is equivalent to reducing $B$, and has no effect on the systematics of our model, since $B$ only enters the equations in the combination $Bt$. The solid line on Fig. \ref{fig:clustergrowth} shows the predictions of our analytic model (Equation \ref{eq:mtotal}) with $A_0=7.96\,{\rm M}_{_\odot}^{1.3}$ and $B=1.53\times 10^{-7}\,{\rm yr}^{-1}$ between $10^6\,{\rm yr}$ and $3\times 10^7\,{\rm yr}$. For comparison, the dotted line shows the corresponding Caldwell \& Chang fit,
\begin{eqnarray}\label{EQN:C&CFit}
M_{\rm tot}&=&{\rm M}_{_\odot}\,\left[\frac{t}{10^6\,{\rm yr}}\right]^2
\end{eqnarray}
between $3\times 10^6\,{\rm yr}$ and $10^7\,{\rm yr}$, but increased by a factor of two  (i.e. incremented by $\log_{_{10}}\!(2)\!=\!0.301$) to avoid confusion. Modulo this increment, the difference between the two fits is less than $10\%$ at all points, and hence less than the Poisson uncertainty due to small-number statistics. In other words, even if the ages and masses of the stars informing the Caldwell and Chang fit were exact, the correspondence would be excellent. We conclude that our model is consistent with the data of Caldwell \& Chang (2018).

%%%%%
\subsection{Caveats}\label{sec:application}
%%%%%

By construction, the stochastic model developed here invokes a balance between the creation of low-mass seed proto-systems by turbulent fragmentation, and their subsequent growth by competitive accretion. This balance is required to maintain an approximately  constant and universal slope, $\alpha\!\sim\! 2.3$, at the high-mass end of the SMF in a forming cluster. Here we review three putative physical processes that might corrupt this balance in nature.

\citet{KrumholzMcKee2008} suggest that there exists a critical column density, $\Sigma_{_{\rm CRIT}}\!\sim\!1\,{\rm g\, cm^{-3}}$, and that at column-densities $\Sigma\!\ga\!\Sigma_{_{\rm CRIT}}$  the radiation from young stellar objects and protostars is trapped and heats the cloud, raising the Jeans mass and thereby promoting the formation of high-mass protostars. The simulations of \citet{Krumholz_etal_2011} suggest that the heating that occurs at high surface density actually has two effects. Not only does it increase the Jeans mass in the vicinity of existing protostars, thereby inhibiting further fragmentation, but it also increases the density at which fragmentation occurs, and hence the accretion rate onto existing protostars. The result is a top-heavy local SMF, with a much higher proportion of high-mass proto-systems than the standard SMF. However, we note that simulations which include the dynamic effects of jets and winds from protostars show that their radiation can easily escape. This keeps the temperature, and hence the Jeans mass, low, and results in a more standard SMF \citep{Krumholz_etal_2011}.

\citet{Li_etal_2010} have explored the effect of the magnetic field strength on the SMF, using ideal MHD simulations of turbulent clouds. They find that the high-mass end of the SMF becomes steeper as the field strength is increased, i.e. the SMF has a lower proportion of high-mass proto-systems than the standard SMF. This is because the gas condenses into filaments aligned perpendicular to the field, and the filaments then fragment into cores which collapse into proto-systems. The proto-systems cannot grow much because they can only accrete along the field, from the diffuse gas outside the filament.

In the ``fragmentation induced starvation'' process reported by \citet{Peters_etal_2010} and \citet{Girichidis_etal_2012}, the disc around a massive protostar fragments to produce a population of low-mass protostars, which promptly consume the remaining disk material, and also intercept any material that flows in subsequently. This process is the {\em opposite} of competitive accretion, since the lower mass protostars grow at the expense of the massive protostar, by cutting it off from its accretion reservoir. However, the massive protostar and its attendant low-mass protostars are -- until and unless they disperse -- a single proto-system, so this process does not change the SMF.

%%%%%
\section{Non-steady cluster growth}\label{SEC:nonsteady}
%%%%%

In our stochastic model (Section \ref{sec:sens}), we fix the fraction of mass going to create new low-mass seed proto-systems, $f_{_{\rm TF}}$. In other words, the mass going to form a star cluster is, {\it throughout the process}, divided in a constant ratio ($f_{_{\rm TF}}:[1-f_{_{\rm TF}}]$) between mass which is consumed in the formation of low-mass seed proto-systems by turbulent fragmentation, and mass which is consumed by competitive accretion.

With this single condition, the {\it rate} at which mass is supplied to the growing cluster becomes immaterial. The shape of the SMF (the log-normal peak at low-masses, and the power-law tail at high masses) changes only in the sense that, as more mass is added, the amplitude increases and the maximum mass increases. The rate of supply of mass to the growing cluster can increase, decrease, or stay constant; it can terminate abruptly and then resume equally abruptly.

Specifically, if the rate at which mass is supplied to the cluster is $dM_{\rm tot}/dt$, we can define
\begin{eqnarray}%CHECKED%
B(t)&=&\frac{[\alpha -2]\,m_0^{[\alpha -2]}}{A_0\,[\alpha -1]\,e^{[\alpha -1]{\cal B}(t)}}\,\frac{dM_{\rm tot}}{dt}\,.
\end{eqnarray}
and
\begin{eqnarray}%CHECKED%
{\cal B}(t)&=&\int\limits_{t'=0}^{t'=t}\,B(t')\,dt'\,.
\end{eqnarray}

If we limit consideration to times, $t$, that do not exceed the time-scale on which the constituent proto-systems evolve significantly or the timescale on which the cluster disperses, (say $10\,{\rm Myr}$), then the total mass of the star cluster, $M_{\rm tot}(t)$, the maximum proto-system mass, $m_{\rm max}(t)$, the number of proto-systems, ${\cal N}(t)$, the rate of creation of proto-systems, $d{\cal N}/dt$, the rate at which mass is consumed in the creation of low-mass seed proto-systems by turbulent fragmentation, $dM_{_{\rm TF}}/dt$, and the rate of consumption of mass by competitive accretion, $dM_{_{\rm CA}}/dt$, are given by
\begin{eqnarray}\label{EQN:Mtot.2}%CHECKED%
M_{\rm tot}(t)&\simeq&\frac{A_0\,e^{[\alpha -1]{\cal B}(t)}}{[\alpha -2]\,m_0^{[\alpha -2]}}\,,\\\label{EQN:mMAX.2}%CHECKED%
m_{\rm max}(t)&\simeq&m_0\,e^{{\cal B}(t)}\,,\\%CHECKED%
{\cal N}(t)&\simeq&\frac{A_0\,e^{[\alpha -1]{\cal B}(t)}}{[\alpha -1]\,m_0^{[\alpha -1]}}\,,\\%CHECKED%
\frac{d{\cal N}}{dt}&\simeq&\frac{A_0\,B(t)\,e^{[\alpha -1]{\cal B}(t)}}{m_0^{[\alpha -1]}}\,,\\%CHECKED%
\frac{dM_{_{\rm TF}}}{dt}&\simeq&\frac{A_0\,B(t)\,e^{[\alpha -1]{\cal B}(t)}}{m_0^{[\alpha -2]}}\,,\\\label{EQN:dMTFdt.2}%CHECKED%
\frac{dM_{_{\rm CA}}}{dt}&\simeq&\frac{A_0\,B(t)\,e^{[\alpha -1]{\cal B}(t)}}{[\alpha -2]\,m_0^{[\alpha -2]}}\,.
\end{eqnarray}

Consequently, $M_{\rm tot}$ can be used in place of $t$ to track the evolution of the global properties of the star cluster:
\begin{eqnarray}%CHECKED%
{\cal N}(t)&\simeq&\frac{[\alpha -2]\,M_{\rm tot}}{[\alpha -1]\,m_0}\,,\\\label{EQN:mMAX.3}%CHECKED%
m_{\rm max}(t)&\simeq&\left\{\frac{[\alpha -2]\,m_0^{[2\alpha -3]}\,M_{\rm tot}}{A_0}\right\}^{1/[\alpha -1]}.
\end{eqnarray}

The use of approximate equalities ($\simeq$) in Equations \ref{EQN:Mtot.2} through \ref{EQN:mMAX.3} reflects the stochastic selection of low-mass seed proto-systems from a log-normal distribution function (see Section \ref{SEC:TurbFrag}). This leads to some noise and the creation of a small number of exceptionally massive proto-systems.

We note that in our model the mass of a cluster is limited by the availability of mass to form new seed proto-systems. Growth of the cluster ceases either because the mass of the proto-cluster cloud is exhausted, or because feedback -- particularly from the most massive proto-systems -- disperses the remaining unaccreted gas. The mass of the most massive proto-system increases monotonically with time (see Equations \ref{EQN:mMAX.1}, \ref{EQN:mMAX.2} and \ref{EQN:mMAX.3}). However, the ratio of mass in any two well populated mass intervals below the current maximum mass does not change significantly with time.

%%%%%
\section{Summary}\label{sec:summary}
%%%%%

We have presented a new phenomenological model for the formation of a star cluster, which can reproduce the Chabrier (2003) System Mass Function very accurately. 

In this model turbulent fragmentation creates low-mass seed proto-systems with a tightly constrained log-normal mass distribution,
\begin{eqnarray}
\frac{dP}{d\mu}&=&\frac{1}{[2\pi ]^{1/2}[0.47]}\;\exp\!\left\{\!\frac{-\,[\mu+0.975]^2}{2\,[0.47]^2}\!\right\},
\end{eqnarray}
where $\mu=\log_{_{10}}\!\!\left(\!m/{\rm M}_{_\odot}\!\right)$.

These proto-systems then grow by competitive accretion. Throughout the formation of the star cluster, a constant fraction $f_{_{\rm TF}}$ of the available mass is consumed by turbulent fragmentation, producing low-mass proto-systems. The remaining $[1\!-\!f_{_{\rm TF}}]$ is consumed by accretion onto these proto-systems. The accretion rate onto an individual proto-system is given by $dm/dt\propto m^\beta$.

If $f_{_{\rm TF}}\!=\!0.23$ and $\beta\!=\!1$, the high-mass tail of the mass function immediately relaxes to the Salpeter slope, and -- in accordance with observations -- retains this slope as the mass of the cluster grows and the mass function extends to ever higher masses. Moreover this behaviour is completely independent of the rate at which mass is supplied to the forming star cluster, even if this rate varies wildly.
These features are still retained approximately if these constraints are relaxed, viz. $0.25\!\la\!f_{_{\rm TF}}\!\la\!0.50$ and $2/3\!\la\!\beta\!\la\!1$.  

We stress that the IMFs/SMFs that emerge from numerical simulations of star formation must be following the model that we present here, since these simulations exhibit competitive accretion onto a growing populations of stars / systems. The division of mass between turbulent fragmentation and competitive accretion probably reflects the fact that the initial condensation of a gravitationally unstable core is highly non-homologous. Consequently the material in the outer envelope is easily stripped away from the much denser central proto-system by ram-pressure or tidal forces, and once this has happened other proto-systems can compete for it. Indeed Hennebelle, Lee \& Chabrier (2019) have argued that the tidal stripping of the outer layers of cores is critical in setting the peak of the IMF. 

The required mass-dependence of the accretion rate appears at first sight to favour tidal-lobe accretion ($\beta\!\simeq\!2/3$) over Bondi-Hoyle accretion ($\beta\!\simeq\!2$). However, there is some evidence from numerical simulations to suggest that high-mass proto-systems are preferentially located in regions where the gas density is lower and/or the velocity dispersion is higher, so that Bondi-Hoyle accretion actually delivers $\beta\!\simeq\!1$.

Evidently further work is needed to explore the physics of turbulent fragmentation, core condensation and competitive accretion. This paper simply outlines a framework that reconciles the notion of turbulent fragmentation with the notion of competitive accretion.

%%%%%
\section*{Acknowledgments}
%%%%%

The authors are indebted to Ralf Klessen, Javier Ballesteros-Paredes, Enrique V{\'a}zquez-Semadeni, Ian Bonnell, Matthew Bate, and Patrick Hennebelle, for enlightening discussions on the nature of the SMF and IMF.  PCC and APW gratefully acknowledge the support of a STFC Consolidated Grant (ST/K00926/1). PCC acknowledges support from the StarFormMapper project, funded by the European Union's Horizon 2020 research and innovation programme under grant agreement No 687528.

%%%%%

%%%%%

%%%%%
\onecolumn
\appendix
\section{Details of the analytic model}
%%%%%

We assume that star formation proceeds by the formation of low-mass seed proto-systems with mass $m_{_{\rm 0}}$, and that these then grow by accretion at a rate
\begin{equation}%CHECKED%
\frac{dm}{dt}=B\,m^{\beta}
\end{equation}
to form higher-mass proto-systems. It follows that the second time derivative of the mass is given by, 
\begin{equation}%CHECKED%
\frac{d^2m}{dt^2}=\beta\,B^2\,m^{[2\beta -1]}\,.
\end{equation}
We also assume that the high-mass end of the proto-system mass function (SMF) in a star forming region is given by
\begin{equation}
\label{EQN:dNdM.t}%CHECKED%
\left.\frac{\partial{N}}{\partial m}\right|_t=A(t)\,m^{-\alpha}\,,
\end{equation}
and we are looking for solutions in which the exponent, $\alpha$, does not change with time. In analysing the high-mass end of the SMF, we start by neglecting the source term for low-mass seed proto-systems. Then we formulate the rate at which low-mass seed proto-systems must be created in order to maintain the resulting SMF. As in the main text, we use standard brackets (e.g. $A(t)$ in Equation \ref{EQN:dNdM.t}) exclusively to denote functional dependence. 

At time $\,t\,$ we identify the cohort of proto-systems in the small but finite mass interval $\;[m(t),m(t)+\Delta m(t)]$. Their number is given by
\begin{eqnarray}\nonumber%CHECKED%
\Delta{N}(t)&=&A(t)\left\{ m^{-\alpha}(t)\,\Delta m(t)\,-\,\frac{\alpha m^{[-\alpha-1]}(t)\Delta m^2(t)}{2}\,+\,{\cal O}\left(\Delta m^3\right) \right\} \\\label{EQN:DeltaN_initial}
&=&A(t)\,m^{-\alpha}(t)\,\Delta m(t)\left\{ 1-\frac{\alpha\Delta m(t)}{2m(t)}+{\cal O}\left(\Delta m^2\right) \right\}.
\end{eqnarray}
By time $\,t+\Delta t\,$, where $\Delta t$ is a small but finite time interval, the lowest mass in this cohort has become
\begin{eqnarray}%CHECKED%
\label{EQN:Mlower}
m(t+\Delta t)\!&\!=\!&\!m(t)\,+\,Bm^\beta(t) \Delta t\,+\,\frac{\beta B^2m^{[2\beta -1]}(t)\Delta t^2}{2}\,+\,{\cal O}\left(\Delta t^3\right)
\end{eqnarray}
Similarly, by time $\,t+\Delta t\,$ the highest mass in this cohort has become
\begin{eqnarray}\nonumber%CHECKED%
m(t+\Delta t)+\Delta m(t+\Delta t)\!&\!=\!&\!\left[m(t)\!+\!\Delta m(t)\right]\,+\,B\left[m(t)\!+\!\Delta m(t)\right]^\beta\Delta t +\;\frac{\beta B^2\left[m(t)\!+\!\Delta m(t)\right]^{[2\beta -1]}\Delta t^2}{2}\,+\,{\cal O}\left(\Delta t^3\right) \\\nonumber
&\!=\!&m(t)\,+\,\Delta m(t)\,+\,Bm^\beta(t)\Delta t\,+\,\beta Bm^{[\beta-1]}(t)\Delta m(t)\Delta t \hspace{1.4cm}\\\label{EQN:Mupper}
&&\hspace{4.2cm}+\,\frac{\beta B^2m^{[2\beta -1]}(t)\Delta t^2}{2}\,+\,{\cal O}\left(\Delta m^2\Delta t,\Delta m\Delta t^2,\Delta t^3\right)\!.
\end{eqnarray}
The mass interval now occupied by the cohort is obtained by subtracting Equation \ref{EQN:Mlower} from Equation \ref{EQN:Mupper} to obtain
\begin{eqnarray}\label{EQN:DeltaMt+Deltat}%CHECKED%
\Delta m(t+\Delta t)&=&\Delta m(t)\,+\,\beta Bm^{[\beta -1]}(t)\Delta m(t)\Delta t\,+\,{\cal O}\left(\Delta m^2\Delta t,\Delta m\Delta t^2\right)\end{eqnarray}

By analogy with Equation \ref{EQN:DeltaN_initial}, the number of proto-systems in the cohort at time $\,t+\Delta t\,$ is given by 
\begin{eqnarray}\label{EQN:DeltaNt+Deltat}%CHECKED%
\Delta {N}(t+\Delta t)\!&\!=\!&\!A(t+\Delta t)\,m^{-\alpha}(t+\Delta t)\,\Delta m(t+\Delta t) \left\{ 1-\frac{\alpha\Delta m(t+\Delta t)}{2m(t+\Delta t)}+{\cal O}\left(\Delta m^2\right) \right\} \! .
\end{eqnarray}
If we focus on zeroth- and first-order terms in Equation \ref{EQN:DeltaNt+Deltat}, we can substitute
\begin{eqnarray}\nonumber%CHECKED%
A(t+\Delta t)&=&A(t)\,+\,\frac{dA}{dt}\Delta t\,+\,{\cal O}\left(\Delta t^2\right)\\
&=&A(t)\left\{ 1\,+\,\frac{dA}{dt}\frac{\Delta t}{A(t)}\,+\,{\cal O}\left(\Delta t^2\right) \right\}\,;\\\nonumber%CHECKED%
m^{-\alpha}(t+\Delta t)&=&m^{-\alpha}(t)\left\{ 1+Bm^{[\beta-1]}(t)\Delta t+{\cal O}\left(\Delta t^2\right) \right\} ^{-\alpha}\\
&=&m^{-\alpha}(t)\left\{ 1-\alpha Bm^{[\beta-1]}(t)\Delta t+{\cal O}\left(\Delta t^2\right)\right\} ,
\end{eqnarray}
where we have used Equation \ref{EQN:Mlower};
\begin{eqnarray}\nonumber%CHECKED%
\Delta m(t+\Delta t)&=&\Delta m(t)\; \left\{ 1^{\,}_{\,}\!+\beta Bm^{[\beta -1]}(t)\Delta t\,+\,{\cal O}\left(\Delta m\Delta t,\Delta t^2\right) \right\},
\end{eqnarray}
where we have used Equation \ref{EQN:DeltaMt+Deltat}; and
\begin{eqnarray}%CHECKED%
\left\{1-\frac{\alpha\Delta m(t+\Delta t)}{2m(t+\Delta t)}+{\cal O}\left(\Delta m^2\right) \right\}&=&\left\{ 1-\frac{\alpha\Delta m(t)}{2m(t)}+{\cal O}\left(\Delta m^2,\Delta m\Delta t\right)\right\} \,.
\end{eqnarray}
Equation \ref{EQN:DeltaNt+Deltat} then becomes
\begin{eqnarray}\nonumber%CHECKED%
\Delta N(t+\Delta t)&=&A(t)\left\{ 1+\frac{dA}{dt}\frac{\Delta t}{A(t)}+{\cal O}\left(\Delta t^2\right)\right\}\hspace{7.8cm} \\\nonumber
&&\hspace{0.8cm}\times\,m^{-\alpha}(t)\!\left\{ 1-\alpha Bm^{[\beta-1]}(t)\Delta t+{\cal O}\left(\Delta t^2\right) \right\} \hspace{0.8cm} \\\nonumber
&&\hspace{1.5cm}\times\,\Delta m(t)\!\left\{ 1+\beta Bm^{[\beta -1]}(t)\Delta t+{\cal O}\left(\Delta m\Delta t,\Delta t^2\right)\right\}  \\\nonumber
&&\hspace{2.5cm}\times\,\left\{ 1-\frac{\alpha\Delta m(t)}{2m(t)}+{\cal O}\left(\Delta m^2,\Delta m\Delta t\right)\right\}  \\\nonumber
&=&A(t)\,m^{-\alpha}(t)\,\Delta m(t)\left\{1+\left[\frac{1}{A(t)}\frac{dA}{dt}-\alpha Bm^{[\beta-1]}(t)+\beta Bm^{[\beta -1]}(t)\right]\Delta t\right.\\\label{EQN:DeltaN_final}%CHECKED%
&&\hspace{5.4cm}\left.\vphantom{\frac{1}{A(t)}\frac{dA}{dt}}-\frac{\alpha\Delta m(t)}{2m(t)}+\,{\cal O}\left(\Delta m^2,\Delta m\Delta t,\Delta t^2\right)\right\} \!,
\end{eqnarray}
The number of proto-systems in the cohort does not change, i.e. $\Delta N(t)=\Delta N (t+\Delta t)$, so we can equate Equation  \ref{EQN:DeltaN_initial} and Equation \ref{EQN:DeltaN_final}. It follows that, in the limit of decreasing $\Delta t$, the coefficients of $\Delta t$ must vanish, so
\begin{eqnarray}%CHECKED%
\frac{1}{A(t)}\frac{dA}{dt}&=&[\alpha-\beta]\,B\,m^{[\beta -1]}(t)\,.
\end{eqnarray}
Since $A$ can not depend on $m$, we must have $\beta =1$, i.e.
\begin{eqnarray}%CHECKED%
\frac{dm}{dt}&=&Bm\,,
\end{eqnarray}
and therefore
\begin{eqnarray}%CHECKED%
A(t)&=&A_0{\rm e}^{[\alpha -1]Bt}\,.
\end{eqnarray}

%%%%%
\end{document}